The existence of precise particle trajectories in any quantum state is accounted for in a consistent way by allowing delocalization of the particle charge. The relativistic mass of the particle remains within a small volume surrounding a singularity moving along the particle trajectory. The singularity is the source of an electric displacement field. The field induces a polarization charge in the vacuum and this charge is equated with the charge of the particle. Under dynamic conditions a distributed charge density $\rho(x,t)$ is induced in the vacuum. The volume integral of the charge density is equal to the charge of the particle and is rigorously conserved. The charge density is derived from a complex-valued physical field $\psi(x,t)$ such that $\rho(x,t) = |\psi(x,t)|^2$. The position probability density is equated with the mean charge density. The mean field $\psi(x,t)$ for many sample realizations with a given energy $E$ and potential $V(x)$ is the sum of the individual fields. In order for the sum to be non-zero, the components in the spectral decompositions of the individual fields must be spatially coherent. The particle has a spin frequency given by Planck's relation $h\nu = T - V + m_0 c^2$, where $T$ is the kinetic energy, determined from the momentum $p$ and $V$ is a quantum potential such that $E = T + V$ is conserved. The instantaneous phase of the spin is given by the phase of exp$(ikx)$ in the spectral decomposition $a(k)$ of $\psi(x,t)$. It is spatially coherent, due to the dependence on $x$. The momentum probability distribution is given by the squared magnitude of the coefficients $a(k)$. The Schrodinger equation is derived by requiring local conservation of mean energy.
\\

# Quantum interference and particle trajectories

**John Luscombe**
*New College, Oxford, OX1 3BN, UK*

The existence of precise particle trajectories in any quantum state is accounted for in a consistent way by allowing delocalization of the particle charge. The relativistic mass of the particle remains within a small volume surrounding a singularity moving along the particle trajectory. The singularity is the source of an electric displacement field. The field induces a polarization charge in the vacuum and this charge is equated with the charge of the particle. Under dynamic conditions a distributed charge density *ρ(x,t)* is induced in the vacuum. The volume integral of the charge density is equal to the charge of the particle and is rigorously conserved. The charge density is derived from a complex-valued physical field *ψ(x,t)* such that $\rho(x,t) = |\psi(x,t)|^2$. The position probability density is equated with the mean charge density. The mean field *ψ(x,t)* for many sample realizations with a given energy *E* and potential *V(x)* is the sum of the individual fields. In order for the sum to be non-zero, the components in the spectral decompositions of the individual fields must be spatially coherent. The particle has a spin frequency given by Planck's relation $h\nu = T - V + m_0 c^2$, where *T* is the kinetic energy, determined from the momentum *p* and *V* is a quantum potential such that $E = T + V$ is conserved. The instantaneous phase of the spin is given by the phase of exp*(ikx)* in the spectral decomposition *a(k)* of *ψ(x,t)*. It is spatially coherent, due to the dependence on *x*. The momentum probability distribution is given by the squared magnitude of the coefficients *a(k)*. The Schrodinger equation is derived by requiring local conservation of mean energy.

## 1. Introduction

According to the principle of complementarity, it is not possible to attribute wave and particle properties to a dynamical system in the same experiment. The principle leads to the conclusion that particles do not follow precise trajectories, since in many experiments, only the wavelike properties can be observed [1].

The double slit experiment provides an example. An interference pattern is produced by illuminating a double slit with a beam of particles. The pattern persists even when one particle at a time passes through the apparatus, provided a large number of particles are accumulated over time.

The particle-like properties can be observed by placing detectors close to the slits. It is found that particles always pass through one slit or the other, but the measurements destroy the interference pattern. If an experiment is designed to measure the interference pattern, then it is not possible to determine which slit the particle passed through. Variations of the experiment are possible, in which the type of measurement is selected randomly at the last moment [2]. The principle of complementarity leads to the conclusion that the trajectory of the particle through the slit, if it has one, is determined only when the particle reaches the measuring apparatus on the other side of the slit.

The deBroglie-Bohm quantum theory of motion denies the validity of the complementarity principle. The basic postulate of Bohm's theory is that particles do follow precise trajectories in space. The motion of the particle is never separated from a quantum field and is fundamentally influenced by the field. Bohm defines a quantum potential, which is treated as a real physical field [3].

Bohm accepts the existence of non-local interactions in quantum mechanics. These are inferred from measurements on pairs of space-like separated particles that have interacted in the past. The experiments show that one measurement event can influence another event when the two events are space-like separated, apparently implying the existence of non-local interactions [4].

The theory discussed in this paper is a development of Bohm's work but non-local interactions are avoided by allowing causal effects to propagate in the backward light cone [5].

The electron is considered to be a relativistic point particle with mass, charge and spin angular momentum. A unified description of the particle and field is provided by treating the particle as a singularity in the electromagnetic field. The relativistic energy of the particle exists as electromagnetic energy, localized at the singularity.

An elementary particle is postulated to have a spin whose frequency is given by Planck's relation

$$v = E / h \tag{1}$$

where $E$ is the total relativistic energy of the particle. The spin frequency defines a radius $r$ at which the linear velocity is equal to the speed of light $c$.

$$r = \frac{\hbar}{mc}$$

The energy of the particle is confined within a sphere of radius $r$, centred at the singularity. Calculation of the angular momentum gives $1/2\hbar$ per axis and a rotational energy equal to $3/4E$. The remaining part of the energy is attributed to the binding energy of the charge. The total relativistic energy can be equated with the energy stored in the field if the coupling strength increases by a factor given by the fine structure constant at distances comparable to $r$ ($10^{-13}$m).

For a particle in a potential well, the spin frequency is given by

$$hv = T - V + m_0 c^2$$

where $T$ is the kinetic energy and $V$ is a quantum potential such that the energy $E = T + V$ is constant. The total phase along a trajectory is given by the complex number

$$\phi = \exp\left(2\pi i \int v dt\right)$$

$$= \exp(iS/\hbar)\exp(im_0 c^2 t/\hbar)$$

where $S$ is the action [6]. The difference in action between two trajectories with the same end points is therefore equal to the difference in total phase.

The phase along any trajectory in the ensemble is postulated to have spatial coherence. If the particle has a momentum $p$ and a position $x$ at time $t$, then the phase is given by the phase of waves existing in the vacuum such that

$$\phi = \exp(ipx/\hbar)\exp(i\omega t)\exp(i\delta)$$

where

$$\hbar\omega = m_0 c^2 - E$$

The rate of change of phase is equal to $v$. The spatial coherence is expressed by requiring that the phase constant $\delta$ be the same for all trajectories in the ensemble. For this condition to hold, a shift in the phase of the spin must occur whenever a stochastic force field existing in the vacuum causes the momentum to change. If this shift is always positive, the total phase will increase less rapidly in regions of space where the waves have similar phases.

Consider the ensemble of all possible particle trajectories with a given momentum distribution. They can be ordered by total phase. The lower the total phase of a trajectory, the more likely that it will occur. The position probability density for the ensemble is determined from the subset of trajectories approaching the minimum value, given by the mean of $S$ for the momentum distribution.

The requirement that the physically occurring trajectories have total phase approaching the minimum value is an expression of the principle of least action. The source of the stochastic field is taken to be random shifts in the phase of the vacuum waves.

A mathematical development of this model is given below.

## 2. Model of the Electron

Consider an electron in its rest frame. According to (1) above, and using the relativistic connection between mass and energy, the spin frequency is given by

$$v_0 = m_0 c^2 / h \qquad (2)$$

where $m_0$ is the rest mass of the particle. If the particle is moving with a velocity $v$ relative to reference frame $S$, then the mass $m$, measured by an observer in $S$, is given by

$$m = m_0 / \sqrt{1 - v^2/c^2}$$

Lorentz invariance requires that the spin frequency in the moving frame is given by

$$v = mc^2 / h \qquad (3)$$

Hence

$$v = v_0 / \sqrt{1 - v^2/c^2}$$

It follows from the principle of relativity that no part of the rotating energy distribution of the particle can be travelling at faster than the speed of light. Hence the maximum radius $r$ of the particle can be taken to be the distance at which the linear velocity is equal to the speed of light. Then

$$r = c/\omega \tag{4}$$

where $\omega$ is the angular frequency given by

$$\omega = 2\pi\nu = mc^2/\hbar$$

Substituting in (4) gives

$$r = \frac{\hbar}{mc} \tag{5}$$

It follows that the radius $r$ undergoes a Lorentz contraction during transformations to inertial frames moving relative to the reference frame.

The energy can be expressed in terms of $r$ to give

$$E = \frac{\hbar c}{r} \tag{6}$$

The value of $r$ for an electron is

$$r = 3.86 \times 10^{-13} m$$

where $\hbar = 1.055 \times 10^{-34} J\sec$, $c = 2.998 \times 10^8 m/s$ and $m = 9.109 \times 10^{-31} kg$.

For a spherically symmetric mass distribution, the moment of inertia about any axis is given by

$$I = \frac{1}{2}mr^2$$

where $m$ is the relativistic mass and the factor *1/2* is a consequence of the inferred radial mass distribution.

The angular momentum about an arbitrary *z*-axis is given by

$$L_z = I\omega$$

Substituting for $I$ and $\omega$ and using (5) gives

$$L_z = 1/2\,\hbar$$

in accordance with the observation that the spin angular momentum of an electron is $1/2\hbar$ per axis. The spin angular momentum is invariant under Lorentz transformations.

The rotational energy about any axis is

$$E = \frac{1}{2}I\omega^2$$

Substituting for $I$ and $\omega$ and using (5) gives

$$E = \frac{1}{4}mc^2 \tag{7}$$

The total rotational energy $U_s$ for all three axes is therefore

$$U_s = \frac{3}{4}mc^2 \tag{8}$$

The remaining part of the relativistic energy is assumed to be due to the binding energy $U_q$ of the electron charge. Hence

$$U_q = \frac{1}{4}mc^2$$

The potential energy of a charge $q$ uniformly distributed over a sphere of radius $r$ is given by

$$U = \frac{1}{2}\frac{q^2}{4\pi\varepsilon_0 r}$$

where $\varepsilon_0 = 8.854 \times 10^{-12}\, Fm^{-1}$ is the permittivity of free space and $q = 1.602 \times 10^{-19}\, C$ is the charge of the electron. This expression can be derived from the capacitance of an isolated sphere $C = 4\pi\varepsilon_0 r$. The stored energy is given by $\frac{1}{2}q^2/C$.

This calculation does not take into account the mutual repulsion of the elements of charge on the sphere. If this is calculated, the result is cut-off dependent and depends on the size of the charge elements. The total stored energy can be written as

$$U = \frac{1}{2}\frac{kq^2}{4\pi\varepsilon_0 r}$$

where $k$ is a measure of the binding strength. $k$ determines the effective granularity of the charge. Equating $U$ and $U_q$ and substituting for $r$ gives

$$k = \frac{1}{2f} \tag{9}$$

where

$$f = \frac{q^2}{4\pi\varepsilon_0 \hbar c} \approx 1/137 \tag{10}$$

is the fine structure constant. In a similar way, the rotational energy $U_s$ can be equated with the stored magnetic field energy of the electron. Calculation gives a coupling strength of

$$k' = \frac{\pi}{2f}$$

This differs by a factor of $\pi$ from equation (9) for the electrostatic case.

A charge $q$ circulating in a loop contributes a current of magnitude

$$i = qv \tag{11}$$

where $v$ is the rotation frequency. The magnetic dipole moment **m** due to the circulating charge is

$$\mathbf{m} = iA\mathbf{n}$$

where $A = \pi r^2$ is the area of the orbit and **n** is a unit vector in the direction of the axis of rotation.. Substituting for $v$ and $r$ gives

$$|\mathbf{m}| = \frac{q\hbar}{2m}$$

which is the usual expression for the $z$-component of the intrinsic magnetic moment of the electron.

The connection between the energy and radius of a particle is given by equation (6),

$$E = \frac{\hbar c}{r}$$

Substitution using equation (10) gives

$$E = \frac{1}{f} \frac{q^2}{4\pi\varepsilon_0 r}$$

If the particle energy is equated with the energy stored in the electromagnetic field, then it follows that $k$ depends on $r$. For large distances $k = 1$, but tends to $1/2f$ at distances of the order of $10^{-13} m$.

The charge of the particle can also be considered as existing in the field. Suppose that the particle is a point source for an electric displacement field **D** such that

$$q = \int_S \mathbf{D}.\mathrm{d}\mathbf{S} \tag{13}$$

where $q$ is the particle charge and the integration is carried out over a surface $S$ enclosing the particle.

Since the vacuum is a polarizable medium, the field induces a polarization **P**, given by **P** = **D**. The potential $V$ due to a polarized dielectric can be attributed to an apparent volume distribution of charge div(**P**) and an apparent surface charge. The potential due to the surface charge is

$$V = \frac{1}{4\pi\varepsilon_0} \int \frac{1}{r} \mathbf{P}.\mathrm{d}\mathbf{S} \tag{14}$$

The contribution from the volume charge is zero, since div**D** = 0 except at the singularity. Evaluating (14) for a spherical surface with constant $r$ and using (13) gives

$$V = \frac{q}{4\pi\varepsilon_0 r}$$

which is the potential at a distance $r$ from a charge $q$.

This shows that the charge of a point source can be treated as equivalent to an induced polarization charge on a small sphere enclosing the particle. The particle charge can be considered as polarization charge in the vacuum and the presence of a charged particle is equivalent to a charge distribution in the field.

In summary, the model treats the electron as a singularity in the electromagnetic field. The singularity is the source of an electric displacement field which induces a charge density in the vacuum. The mass-energy of the particle is equated with the energy stored in the field of the singularity.

## 3. Motion of a Particle in a Potential Well

### 3.1 Kinetic and Potential Energy

The spin frequency of a particle is given by (3) above. The kinetic energy $T$ is defined by

$$T = mc^2 - m_0 c^2$$

Substituting in (3) gives

$$v = (T + m_0 c^2)/h \tag{15}$$

Suppose that a particle, initially at rest, falls into a potential well and radiates energy. The initial energy of the system is just the rest masses of the particle and the potential. The total energy after capture of the particle is given by

$$E_t = m_0 c^2 + m_1 c^2 + E$$

where $m_1$ is the rest mass of the potential and $E$ is the (negative) radiated energy. The total energy for the motion of the particle is equal to $E$. It is the sum of the terms

$$E = T + V \tag{16}$$

where $T$ is the kinetic energy of the particle and $V$ is the potential energy stored in the field.

The potential energy has a rest mass given by $V/c^2$. Since $V$ is always negative in a potential well, the associated rest mass is negative.

In order that the combined system of the potential well and the particle transform correctly under Lorentz transformations, the combined rest mass of the particle and the energy stored in the potential must be equal to $m_0$. Consequently, the rest mass of the particle is reduced by $V/c^2$.

Equation (3) then becomes

$$v = (T - V + m_0 c^2)/h \tag{17}$$

Substitution of (16) into (17) gives

$$v = (2T - E + m_0 c^2)/h \tag{18}$$

The physically observable properties of the motion of the particle are position and momentum probability distributions for an ensemble of identically prepared physical systems. In the following description, the dynamical variables are treated as sets of random variables, indexed by the time *t*. They are defined for an ensemble of sample realizations. The particle position for the *n*th trajectory in the ensemble is denoted by $X_n(t)$. The process is assumed to be ergodic, that is to say, ensemble averaging is equivalent to time averaging. Consequently the trajectory of a single particle, averaged over time, has the same statistical properties as an ensemble of trajectories.

The kinetic and potential energies along the *n*th trajectory can be represented by $T_n(t)$ and $V_n(t)$. The potential energy is no longer a deterministic function of position. It is not the classical potential energy but simply the energy stored in the vacuum field. *E* and *V* are always negative quantities and *T* can take any positive value.

The instantaneous spin frequency $v_n$ for the *n*th trajectory is given by equation (18).

$$v_n = \left(2T_n(t) - E + m_0 c^2\right)/h \tag{19}$$

The change of particle kinetic energy with time is accounted for by the action of the vacuum field acting on the particle. The equation of motion is

$$m\dot{v} = \Gamma_n(t)$$

where $\Gamma_n(t)$ represents the stochastic force acting on the particle along the *n*th trajectory.

The total phase along the *n*th trajectory is given by

$$\theta_n = 2\pi \int v_n dt \tag{20}$$

In the special case when no force acts on the particle, the velocity is constant and equation [20] can be integrated to give

$$\theta = \left(p.x - Et + m_0 c^2 t\right)/\hbar \tag{21}$$

where *p* is the constant momentum.

## 3.2 Self Field of the Electron

It is known that the position and momentum distributions for the motion of an electron are correctly predicted by quantum mechanics. Given that the instantaneous spin frequency along the *n*th trajectory is $v_n$, what additional hypothesis is required to give a theory that replicates the predictions of quantum mechanics?

According to the model of the electron developed earlier, the electric displacement field of the particle induces a charge density in the vacuum. For a particle at rest, the charge distribution is a surface charge on a sphere enclosing the source. When the particle is in motion, the charge density becomes a volume distribution of charge.

Let $\rho(x,t)$ denote the instantaneous charge density. The following hypothesis is now introduced: There exists a physical field $\psi(x,t)$, defined in four-dimensional space-time, such that

$$\rho(x,t) = |\psi(x,t)|^2 \qquad (22)$$

Conservation of charge requires that

$$\int_V \rho(x,t) dv = q$$

The value of $\rho(x,t)$ is invariant under local gauge transformations of the type [7]:

$$\psi'(x,t) = e^{i\theta(x)} \psi(x,t)$$

Let $\psi_n(x,t)$ denote the field for *n*th member of an ensemble of trajectories. The charge density is given by equation (22). The instantaneous potential energy $V_n(t)$ is the work done to establish the charge density in the potential $V(x)$.

$$V_n(t) = \int_V V(x) \rho_n(x,t) dv$$

Experimental observations for an ensemble of identically prepared systems give a position probability density $P(x)$ for the particle. The units of charge can be selected so that $q = 1$ and the mean charge density is then defined as equivalent to the position probability density.

The mean charge density is postulated to be derived from a mean field $\psi(x,t)$. The position probability density is given by

$$P(x,t) = |\psi(x,t)|^2$$

with

$$\int_V |\psi(x,t)|^2 dv = 1 \qquad (23)$$

The mean field for an ensemble of trajectories is the sum of the individual fields,

$$\psi(x,t) = C \sum \psi_n(x,t)$$

$C$ is a normalization constant, required for $\psi(x,t)$ to satisfy equation (23).

The instantaneous field $\psi_n(x,t)$ at any time can be expressed as a Fourier transform,

$$\psi_n(x) = \int a_n(k) e^{ikx} dk$$

The sets of coefficients $a_n(k)$ define a vector in a Hilbert space. The summation is therefore a summation of vectors in the Hilbert space. The mean field is given by the direction of the vector, not its magnitude. This allows the normalization constant to be set to any value.

The mean field can also be expressed as a Fourier transform,

$$\psi(x) = \int a(k) e^{ikx} dk$$

where

$$a(k) = C \sum a_n(k)$$

If the $a_n(k)$ have random phase, then the $a(k)$ are zero. This implies that, for non-zero $a(k)$, the fields of the sample trajectories are phase coherent. Since the samples can be widely separated in space and time, there must exist a wave field in the space, with spatial components that have a constant relative phase.

If the wave field has a time dependency $\exp(i2\pi vt)$, the field components are travelling waves. The time dependency does not change the charge distribution and the wave field may be written as

$$\psi(x,t) = e^{i2\pi vt} \int a(k) e^{ikx} dk \qquad (24)$$

The particle may be considered to act as a point source for the field. Suppose that the value of the field along the $n$th trajectory is represented by the complex number

$$\phi_n(t) = \exp(2\pi i \int v_n dt)$$

where $v_n$ is given by equation (19). It is postulated that if a particle in a potential well has a momentum $p$ at position $x$ and time $t$, the instantaneous phase of the spin is given by $\arg(\phi)$, where

$$\phi = \exp\left[i\left(px - Et + m_0 c^2 t\right)/\hbar\right] \tag{25}$$

In other words, the spin has the same phase as a particle travelling with constant momentum at position $x$, given by equation (21).

Consider a small volume surrounding the point $x$. The mean field $\phi(x)$ at $x$ is the sum of the fields $\phi_n(x)$ for the trajectories passing through the volume at time $t$.

Equation (25) allows $\phi(x,t)$ to be expressed as a summation over the momentum distribution

$$\phi(x,t) = \int a(p) \exp(ipx/\hbar) dp \cdot \exp\left[i\left(m_0 c^2 - E\right)t/\hbar\right]$$

provided that the coefficients $a(p)$ are not dependent on position, that is to say the momentum distribution of the particle is the same everywhere in space.

Let $Q(p)$ be the momentum probability density. The last postulate is given by

$$Q(p) = |a(p)|^2 \tag{26}$$

This can be taken to define a charge density in momentum space. The proportion of time spent by the particle with momentum in the interval $p$ to $\Delta p$ is given by $Q(p)\Delta p$. If the charge carried by the vacuum waves in this interval is proportional to the time the momentum lies within the interval, then the charge in the interval can be equated with the square of the wave amplitude $a(p)$, times $\Delta p$.

The mean field $\phi(x,t)$ is derived from a summation for the trajectories. It can be equated with the mean field $\psi(x,t)$, given by equation (24). Then

$$k = \frac{p}{\hbar}$$

$$v = \left(m_0 c^2 - E\right)/h$$

$$\psi(x,t) = \int A(p)dp \cdot \exp[i(p.x - Et + m_0c^2 t)/\hbar] \qquad (27)$$

The wave speeds $w$ of the components in the summation can be determined from the relativistic equation connecting momentum and energy

$$E^2 = c^2 p^2 + m_0^2 c^4$$

$$w = c\sqrt{1 + (m_0 c / p)^2}$$

The wave speed tends to infinity as $p$ tends to zero and it approaches $c$ from above as $p$ becomes very large. Wave speeds greater than the speed of light are consistent with the theory of relativity, provided that it is not possible to construct an experiment in which super-luminal signalling occurs.

### 3.5 Schrodinger Equation

For equation (25), it follows that the instantaneous momentum is given by

$$\frac{\partial \psi}{\partial x} = \frac{ip}{\hbar}\psi$$

The instantaneous kinetic energy $T$ is given by

$$\frac{\partial^2 \psi}{\partial x^2} = \frac{-p^2}{\hbar^2}\psi$$

with $\quad T = \dfrac{p^2}{2m}$

The constant total energy is given by

$$\frac{\partial \psi}{\partial t} = \frac{i}{\hbar}(m_0 c^2 - E) \qquad (28)$$

The mean value of the field, summed over all trajectories is given by equation (27). It follows from equation (26) that

$$\int_V \psi^*(x,t)\left(-i\hbar \frac{\partial}{\partial x}\right)\psi(x,t)dv = \bar{p}$$

where $\bar{p}$ is the mean momentum, defined by

$$\bar{p} = \int_V p|A(p)|^2 dp$$

and (*) denotes complex conjugation. It also follows that the mean kinetic energy is given by

$$\bar{T} = \int_V \psi^*(x,t)\left(\frac{-\hbar^2}{2m}\frac{\partial^2}{\partial x^2}\right)\psi(x,t)dv$$

The mean potential energy at any point $x$ can be considered as the work done to bring a charge $|\psi|^2$ to a potential $V(x)$. The mean potential energy over all space is given by

$$\bar{V} = \int_V \psi^*(x,t)V(x)\psi(x,t)dv$$

From equation (28), the constant total energy of the motion can be written as

$$E = \int_V \psi^*(x,t)\left(i\hbar\frac{\partial}{\partial t} + m_0 c^2\right)\psi(x,t)dv.$$

Conservation of energy requires that

$$E = \bar{T} + \bar{V} \tag{29}$$

One wishes to find a differential equation for $\psi(x,t)$ with this property. The simplest equation is found by requiring that equation (29) hold at all points in space. The terms under the integral signs can then be equated. This gives

$$\frac{-\hbar^2}{2m}\frac{\partial^2}{\partial x^2}\psi(x,t) + \left(V(x) - m_0 c^2\right)\psi(x,t) = i\hbar\frac{\partial}{\partial t}\psi(x,t) \tag{30}$$

When $V = 0$, the equation has solutions:

$$\psi(x,t) = \exp\left[i\left(\pm p.x - Et + m_0 c^2 t\right)/\hbar\right].$$

Equation (30) is the Schrodinger wave equation with an extra term involving the rest mass. This term appears in the Dirac equation and but was not carried through by Schrodinger.

## 4. Conclusion

The theory developed here replicates the observational consequences of the standard theory of quantum mechanics, often known as the Copenhagen interpretation. Consequently, the two theories cannot be distinguished on experimental grounds. The criterion that should be applied instead is the logical consistency of the underlying assumptions.

Unlike the Copenhagen interpretation, the deBroglie-Bohm theory assigns no special role to the observer. The principle of complementarity requires that the wave properties of a particle are not determined until an observer causes an act of measurement. The essential role played by the observer leads to paradoxes relating to the collapse of the wave function. These paradoxes cannot be resolved satisfactorily within the Copenhagen interpretation, whereas the deBroglie-Bohm theory has equal or greater empirical content but is free of paradoxes and can be developed as a local theory.